# Cellular Automata and Its Applications in Bioinformatics: A Review

Pokkuluri Kiran Sree [1], Inampudi Ramesh Babu [2], SSSN Usha Devi .N[3]

1. Research Scholar, Dept of CSE, JNTU Hyderabad, profkiransree@gmail.com

2. Professor, Dept of CSE, Acharya Nagarjuna Univesity, Guntur. drirameshbabu@gmail.com

3. Assistant Professor, Dept of CSE, JNTU Kakinada

*Abstract*

This paper aims at providing a survey on the problems that can be easily addressed by cellular automata in bioinformatics. Some of the authors have proposed algorithms for addressing some problems in bioinformatics but the application of cellular automata in bioinformatics is a virgin field in research. None of the researchers has tried to relate the major problems in bioinformatics and find a common solution. Extensive literature surveys were conducted. We have considered some papers in various journals and conferences for conduct of our research. This paper provides intuition towards relating various problems in bioinformatics logically and tries to attain a common frame work for addressing the same.

## Introduction

The survey is mainly concerted survey of problems in bioinformatics like protein coding and promoter prediction and protein structure prediction. In the section 3 of this paper, Cellular Automata is introduced with AIS-MACA (Artificial Immune System Based Multiple Attractor Cellular Automata). The next section 4, 5 of the paper introduce the survey on protein coding region, promoter region prediction. Then the survey on protein structure prediction is presented in section 6 and section 7 that a common frame work tool output based on CA.

## Cellular Automata

Artificial Intelligence (AI) is the exploration of impersonating human mental ability in a computer.

Current AI methods predominantly succumb to two general categories.

*Explicit Modeling (Words, Images)*

This is based on rules, frames and case based learning has succeeded in some domains. This category cannot handle the unseen cases.

*Implicit Modeling (Numerical Techniques)*

The second methodology addresses this issue through advancement of the model dependent upon perceptions and experience. In accordance with this approach, CA based model takes in companionships from a set of samples and after that apply this information base to handle beforehand unseen in line with this methodology, Cellular Automata (CA) based classifier can learn associations from examples and apply this acquired knowledge to handle unknown cases.

1) *Cellular Automata*

Cellular automata consist of a grid of cells with a finite number of states. Cellular Automata (CA) is a computing model which provides a good platform for performing complex computations with the available local information.

CA is defined a four tuple <G, Z, N, F>

Where   G -> Grid (Set of cells)

    Z -> Set of possible cell states

    N -> Set which describe cells neighborhoods

    F -> Transition Function (Rules of automata)

A CA displays three basic characteristics - locality, infinite parallelism, and simplicity

**Locality:** CA is characterized by local connectivity of its cells. The interactions among the cells are defined on the basis of locality. Each cell can communicate with adjacent cells. The transitions defined between cells carry small amount of information only. None of the cells connected will have a global view of the complex system.

**Parallelism:** Parallel computing environment is demanded for addressing most of the complex computing problems. Most parallel computers contain more than a few dozens of processors. CA can achieve parallelism on a scale larger than massively parallel computers. CA performs computations in a distributed fashion on a spatially extended grid. It differs from the conventional approach to parallel computation in which a problem is split into independent sub-problems, each solved by a different processor; the solution of sub-problems are subsequently combined to yield the final solution

**Simplicity:** The basic unit of Cellular Automata (CA) is a cell that has a simple structure evolving in discrete time and space. Wolfram developed and proposed a uniform cellular automata construction which is a significant contribution in the field of computing. This leads to the emergence of single dimensional array approach with each cell has only two states and three neighborhoods. It is very easy to map the frame work of CA to the frame work of many bioinformatics problems.

As Cellular Automata consists of a number of cells structured in the form of a grid. The transitions between the cells may depend on its own state and the states of its neighboring cells. The equation one sates that if $i^{th}$ cell have to make a transition, it has to depend on own state, left neighbor and right neighbor either.

$q_i(t + 1) = f(q_{i-1}(t), q_i(t), q_{i+1}(t))$    ---- Equation1.1

### 2) Wolfram's two state three neighborhood Cellular Automata

Figure 1, 2 demonstrate Wolfram's two state three neighborhoods, which consist of eight different present state combinations. The rule 254 is used for defining the transitions between the neighbors. Each CA cell in the grid must be a memory element (D Flip flop) with some combinational logic - an XOR and or XNOR Gate (additive). Each cell will be updated at every clock cycle. The transitions between the cells will depend on the immediate neighbors

TABLE1. RULE REPRESENTATION

| Possible Combinations | 111 | 110 | 101 | 100 | 011 | 010 | 001 | 000 |
|---|---|---|---|---|---|---|---|---|
| Binary Equivalent of Rule-254 (Next State) | 1 | 1 | 1 | 1 | 1 | 1 | 1 | 0 |

The decimal equivalent of the next state function is defined as the rule number of the CA cell introduced by Wolfram. In a 2-state 3-neighborhood CA, there are 256 distinct next state functions. Out of 256 rules, rule 254 is demonstrated. In Cellular Automata rules play a vital role in developing a good classifier.

### 3) AIS-MACA (Artificial Immune System Based Multiple Attractor Cellular Automata)

Multiple Attractor Cellular Automata which is a special class of fuzzy cellular automata which was introduced thirty years ago. It uses fuzzy logic to handle real value attributes. The development process / implementation of Multiple Attractor Cellular Automata are administered by AIS technique, a Clonal Algorithm with the underlying theory of survival of the fittest gene. Genetic algorithm framework is adapted to represent the corresponding CA rule structure. So we have named this special classifier as AIS-MACA. AIS-MACA can address many problems in bioinformatics

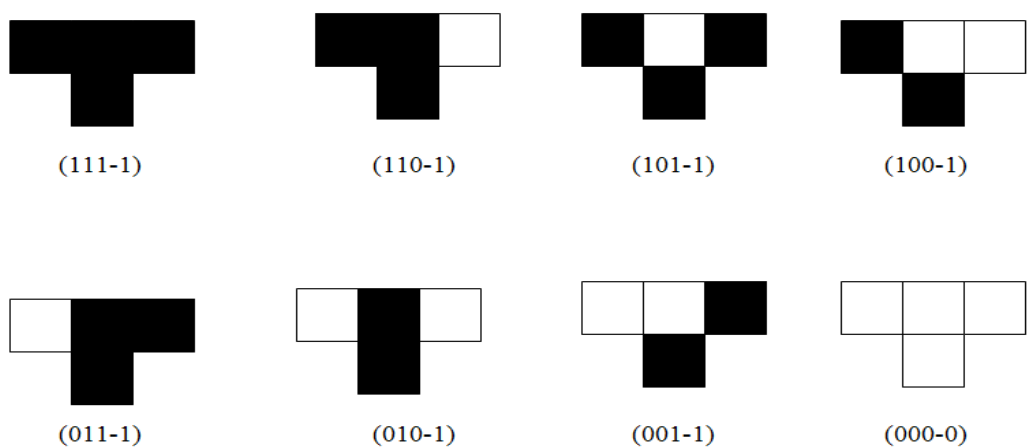

FIGURE 1. NEIGHBORHOOD REPRESENTATION

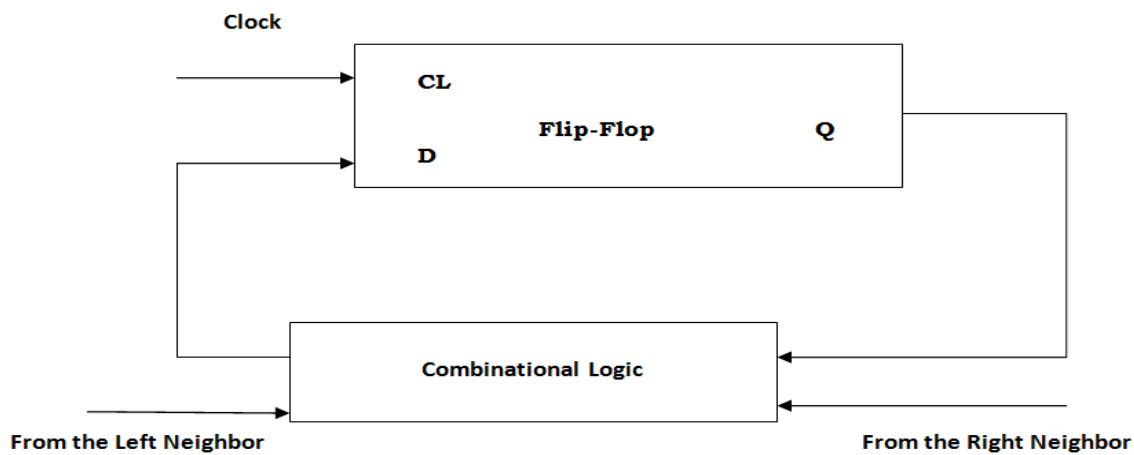

FIGURE 2. WOLFRAM REPRESENTATION OF FLIP-FLOP CIRCUIT

**Protein Coding Regions**

Jesus P. Mena-Chalco at al has used Modified Gabor-Wavelet Transform for addressing this issue. In this connection, numerous coding DNA model-free systems dependent upon the event of particular examples of nucleotides at coding areas have been proposed. Besides, these techniques have not been totally suitable because of their reliance on an observationally predefined window length needed for a nearby dissection of a DNA locale. Authors present a strategy dependent upon a changed Gabor-wavelet transform for the ID of protein coding areas. This novel convert is tuned to examine intermittent sign parts and presents the focal point of being free of the window length. We contrasted the execution of the MGWT and different strategies by utilizing eukaryote information sets. The effects indicate that MGWT beats all evaluated model-autonomous strategies regarding ID exactness. These effects demonstrate that the wellspring of in any event some pieces of the ID lapses handled by the past systems is the altered working scale. The new system stays away from this wellspring of blunders as well as makes an instrument accessible for point by point investigation of the nucleotide event.

Changchuan Yin at al has proposed a strategy to foresee protein coding areas is produced which is dependent upon the way that the vast majority of exon arrangements have a 3-base periodicity, while intron groupings don't have this interesting characteristic. The technique registers the 3-base periodicity and the foundation clamor of the stepwise DNA sections of the target DNA groupings utilizing nucleotide circulations as a part of the three codon positions of the DNA successions. Exon and intron successions might be recognized from patterns of the degree of the 3-base periodicity to the foundation commotion in the DNA groupings.

Suprakash Datta at al used a DFT based gene prediction for addressing this problem. Authors provided theoretical concept of three periodicity properties observed in protein coding regions in genomic DNA. They proposed new criteria for classification based on traditional frequency approaches of coding regions.

Pradipta Maji has proposed a Cellular Automata (CA) based example classifier to recognize the coding locale of a DNA succession. CA is exceptionally straightforward, proficient, and handles more faultless classifiers than that has long ago been acquired for a reach of diverse succession lengths. Far reaching trial outcomes make that the proposed classifier is a financially savvy elective in protein coding area identification issue.

We also proposed a classifier based on MACA for predicting the protein coding region with an accuracy of 83%. It is later strengthened with an AIS technique whose predicted accuracy is more compared to the earlier.

**Promoter Region Prediction**

Anne Vanet at al proposed some algorithms for predicting the promoter regions. This paper displays a study of right now accessible scientific models and algorithmically routines for attempting to recognize promoter successions. The routines concern both looking in a genome for an awhile ago characterized agreement and concentrating an accord from a set of arrangements. Such routines were frequently customized for either eukaryotes or prokaryotes in spite of the fact that this does not block utilization of the same system for both sorts of organic entities. The overview consequently blankets all techniques; in any case, attention is set on prokaryotic promoter arrangement identification. Illustrative requisitions of the fundamental concentrating calculations are given for three microbes

Vetriselvi Rangnanan at al made a dissection of different anticipated structural lands of promoter districts in prokaryotic and in addition eukaryotic genomes had prior shown that they have a few normal characteristics, for example, lower steadiness, higher curve and less bendability, when contrasted and their neighboring areas. In light of the distinction in solidness between neighboring upstream and downstream areas in the region of tentatively dead set translation begin destinations, a promoter forecast calculation has been produced to distinguish prokaryotic promoter groupings in entire genomes

Jih-Wei Hung has developed an effective forecast calculation that can expand the recognition (power =1 - false negative) of promoter. Authors introduce two strategies that utilize the machine force to ascertain all conceivable examples which are the conceivable characteristics of promoters. The primary strategy we exhibit FTSS (Fixed Transcriptional Start Site) utilizes the known TSS positions of promoter arrangements to prepare the score record that helps us in promoter forecast. The other strategy is NTSS (Nonfixed TSS). The TSS positions of promoter arrangements utilized as a part of NTSS are thought to be obscure, and NTSS won't take irrefutably the positions of Tsss into attention. By the exploratory effects, our expectation has higher right rate than different past systems

Marshall S.Z. Horwitz at al have chosen an assembly of Escherichia coli promoters from irregular DNA groupings by swapping 19 base sets at the -35 promoter area of the etracycline safety gene te" of the plasmid pbr322. Substitution of 19 base sets with artificially blended irregular groupings brings about a greatest of 419 (something like 3 x 1011) conceivable swap groupings. From a populace of in the ballpark of 1000 microscopic organisms harboring plasmids with these irregular substitutions, tetracycline choice has uncovered numerous practical -35 promoter successions. These promoters have held just halfway. Homology to the -35 promoter accord grouping. In three of these promoters, the agreement operator moves 10 nucleotides downstream, permitting the RNA polymerase to distinguish an alternate Pribnow box from inside the definitive pbr322 succession. Two of the successions advertise translation more determinedly than the local promoter

We also proposed a classifier based on MACA for predicting the promoter region with an accuracy of 89%. It is later strengthened with an AIS technique whose predicted accuracy is more compared to the earlier

**Protein Structure Prediction**

Jooyoung Lee, Anticipating protein 3d structures from the amino harsh corrosive arrangement still stays as an unsolved issue after five decades' endeavors. In the event that the target protein has a homologue recently understood, the undertaking is moderately simple and high-determination models could be assembled by replicating the skeleton of the settled structure. Be that as it may, such a modeling methodology does not help address the inquiry of how and why a protein embraces its particular structure. In the event that structure homologues (sporadically analogues) don't exist, or exist can't be recognized yet, models must be developed starting with no outside help. This strategy, called stomach muscle initio modeling, is crucial for a complete answer for the protein structure forecast issue; it can likewise help us comprehend the physicochemical rule of how proteins crease in nature. Right now, the precision of abdominal muscle initio modeling is low and the victory is restricted to little proteins (<100 buildups). In this section, we give an audit on the field of stomach muscle initio modeling. Center will be put on three key components of the modeling calculations: vigor capacity, conformational inquiry, and model determination. Advances and developments of an assortment of calculations will be talked about.

Scott C. Schmidler has provided a significant part for statisticians in the period of the Human Genome Project has improved in the rising territory of structural bioinformatics". Succession dissection and structure expectation for biopolymers is a vital venture on the way to transforming recently sequenced genomic information into naturally and pharmaceutically pertinent data in backing of sub-atomic solution. We depict our work on Bayesian models for expectation of protein structure from succession, in view of dissection of a database of tentatively resolved protein structures (Datta, Suprakash et al., 2005)

We have advanced fragment based models of protein optional structure awhile ago which catch central parts of the protein collapsing procedure. These models give prescient executions at the level of the best accessible strategies in the. Here we indicate that this Bayesian schema is regularly summed up to fuse data dependent upon non-neighborhood arrangement cooperation's. We show this thought by displaying a basic model for strand blending and a Markov chain Monte Carlo (MCMC) calculation for induction. We apply the methodology to forecast 3-dimensional contacts for two illustration proteins.

David E. Kim al worked on PSP and the arrangements submitted to the server are parsed into putative dominions and structural models are produced utilizing either near displaying or a new structure expectation routines. Assuming that a certain match to a protein of known structure is found utilizing BLAST, PSI-BLAST, Ffas03 or 3d-Jury, it is utilized as a pattern for relative displaying. Assuming that no match is found, structure expectations are made utilizing the Rosetta piece insertion strategy again. Test atomic attractive thunder (NMR) obligations information can likewise be submitted with an inquiry succession for Rosetta NMR all over again structure determination. Other current capacities might be the expectation of the impacts of transformations on protein–protein communications utilizing computational interface alanine filtering. The Rosetta protein configuration and protein–protein docking systems will soon be accessible through the server.

We also proposed a classifier based on MACA for predicting the secondary structure of protein with an accuracy of 85%. It is later strengthened with an AIS technique whose predicted accuracy is more compared to the earlier.

## CA Based Integrated Tool For Predicting Protein Coding and Promoter Region

We have proposed an integrated tool showed in Fig 3.4 for predicting the protein coding and promoter region using the same classifier. This has recognized an international accload. We then extended this framework to predict the secondary structure of the protein. The classifier is trained with Fickett and Toung data sets for protein coding regions and E Coli data sets for promoter prediction. This proposed algorithm can handle different sequence lengths input and can give an accuracy of 82%.

```
   1  AAGCTTGGCA GCAAAAGGTG CCTGAGGCTA CGTTTTAATT ATTGGCTCCA GAGAGCGGGC CAACGCCTGC ATTGCACAGG   80
  81  ACATGCGGCG GGCAGACACA GTAGCACACA CCTACACGCA CTCGCACACT CAGGTGGGCT TGGTGAGCCT GTCTTCCACG  160
 161  AGGCTCCGCG AACGTCCCAA CCCGTCACAG GCGCGGAACT CCAAGCGCGT GGAGTTCAAG CGGTACTTAC CCCCAAGACG  240
 241  CTCTGAGCCA GGAGCCCCAC TGTCACAGTC GCCAGCAGCA GCCACCTGCA TAGGAGAAGG AGGGAGGGTG AGAGAGCGAC  320
 321  ATAAAAGTGC GGGCCCCAAC TTGTCAACTC TAGAAGATA CATTCAAAGT CCCCTGCGCC TAGGAGACCA GGCAGAAGGC  400
 401  GACTTACCCT CTCAGGGGAA GACCAGAGGC TGAGAACCGC ACTCGGTCCA TGATCCCAGT CTTTCAGCTC TAGGCTCGGC  480
 481  TCGATCCCTT CCTAACCTGG AGGCAAGAGA CAAAGAGGGT TAGCAAGGGT TGCAGGCAGA GAGAGCATTA TCATCGCAGC  560
 561  ACCCTGACTC CGCAGTAATG TCCCTCGTTT CCTTGCGGAG TGACCAAAGT TCTGTTCCAC TAGGAGACAG AAAAAGGGAG  640
 641  ACTCTTCTCC GTAGGCCCTG CCTGGAGCTG TTCAGCCCAT GCAGCGCGAT CCTGATTCCT CCAGCTCCGA TCCAACCGAG  720
 721  CGCTCCGCCT TAGCCTGCCC CAGCGCTACA AGTCCGCGAC TTGGGGCCTG GCTCTCGGCG CCTCCCGGGG GTCTCTCTCA  800
 801  CCTCGGGGTC AGGAGCTGCT GCCGGCGCTC ATGGGCCCCT GCTGAAGCGT TGGCTGCGGA CACACTGGCG GCCAGGCCTG  880
 881  CCTGGCTCCG CTTGGTGGGC GCCGGATTCC CGGCCCCGGC GTGCCTTGCC GGTGCTTGCC CGGGATTGGC GAGCCAAGGG  960
 961  GCTCGTGGAA TCGGGGCGCG GGAAGAGCTG GGTGCGGAGC GCAGAGCCCC GGCGTTCGGA CTCCCTGCGA CACCAAGTCC 1040
1041  CGAGCCGCGC GCAGCCTCCG CCTCTTACCC GCGCCGCAGG GTCCTCCCCT TGGAGACGCC GCCCGCGCAC TGCGGAGGGG 1120
1121  AGGGGGCAGC GCCAACAAAT TGGGGAGCTC CTCGGGCGGC GCGCTCAGGT CTCTGCTCAG AGCCACCGCA CCCGGGACGG 1200
1201  TGAGCAGGGC TCACGCCCGG GCCCTCCGAG ACTGAGCACC TCGCCGCGGC CCACGGGAGA CGGGCTGCAG TTCCCCGCCG 1280
1281  TCTCTGTTGG TCCCCCGCCG CCCCGGGCGC GCCACCATGG GGCCCCGGCT CAGCGTCTGG CTTCTGCTGC CGCTCGCCGC 1360
1361  CCTTCTGCTC CACGAGGAGC GCAGCCGGGC CGCTGCGAAG GTGAGTCCCC TTCGGGCTGC GCTCCCCAAC CTCCTTCCTT 1440
1441  GCACCTCCGG CGCCCCGCGA GGCCTCTTTG TCCAGCGCCT GGCGCGGTCA AGCCGCCCGG CCCTCGAGGG TGTCGCCGGT 1520
1521  GGCACGGCCA GGTGCACTCT TCCACAGGCT CTCCCGCTCG TGGTCCCTGT GGGCTGCGAT CCACATCCA TTACGGAACC 1600
1601  TCCTTTGTTA CGGGCTGGGT GAGGGAGACT TAGGAAGGGA CTGAAGCTTA AGCTTGGCAG CAAAAGGTGC CTGAGGCTAC 1680
1681  GTTTTAATTA TTGGCTCCAG AGAGCGGGCC AACGCCTGCA TTGCACAGGA CATGCGGCGG GCAGACACAG TAGCACACAC 1760
1761  CTACACGCAC TCGCACACTC AGGTGGGCTT GGTGAGCCTG TCTTCCACGA GGCTCCGCGA ACGTCCCAAC CCGTCACAGG 1840
1841  CGCGGAACTC CAAGCGCGTG GAGTTCAAGC GGTACTTACC CCAAGACGC TCTGAGCCAG GAGCCCCACT GTCACAGTCG 1920
1921  CCAGCAGCAG CCACCTGCAT AGGAGAAGGA GGGAGGGTGA GAGAGCGACA TAAAAGTGCG GGCCCCAACT TGTCAACTCT 2000
2001  AGAAAGATAC ATTCAAAGTC CCCTGCGCCT AGGAGACCAG GCAGAAGGCG ACTTACCCTC TCAGGGGAAG ACCAGAGGCT 2080
```

FIGURE 3. COMMON PROTEIN CODING & PROMOTER PREDICTION INTERFACE

| Start Position | End Position | Strand | Score | Sequence |
|---|---|---|---|---|
| 3618 | 3625 | + | 7.86 | CATAAAAG |
| 1969 | 1976 | + | 7.86 | CATAAAAG |
| 320 | 327 | + | 7.86 | CATAAAAG |

FIGURE 4. BOUNDARY REPORTING FOR PROMOTER

## Conclusion

This paper is aimed at bringing an intuition towards application of CA in bioinformatics. A common frame work is required for addressing major problems in bioinformatics (Crick Francis et al., 1970). CA frame work can be used to address many problems in like protein structure prediction, RNA structure prediction, predicting the splicing pattern of any primary transcript and analysis of information content in DNA, RNA, protein sequences and structure and many more.